# 'Q-Feed' - An Effective Solution for the Free-riding Problem in Unstructured P2P Networks

Sabu M. Thampi[†], Chandra Sekaran K[††]
[†]L.B.S Institute of Technology for Women, Kerala-695012, India
[††]National Institute of Technology Karnataka, Surathkal, Karnataka-575025, India
*smtlbs@in.com, kch@nitk.ac.in*

*Abstract-* **This paper presents a solution for reducing the ill effects of free-riders in decentralised unstructured P2P networks. An autonomous replication scheme is proposed to improve the availability and enhance system performance. Q-learning is widely employed in different situations to improve the accuracy in decision making by each peer. Based on the performance of neighbours of a peer, every neighbour is awarded different levels of ranks. At the same time a low-performing node is allowed to improve its rank in different ways. Simulation results show that Q-learning based free riding control mechanism effectively limits the services received by free-riders and also encourages the low-performing neighbours to improve their position. The popular files are autonomously replicated to nodes possessing required parameters. Due to this improvement of quantity of popular files, free riders are given opportunity to lift their position for active participation in the network for sharing files. Q-feed effectively manages queries from free riders and reduces network traffic significantly.**

## I. Introduction

A P2P network serves the content among the associate nodes rather than focussing it at a single central server. The barriers to starting and growing such systems are low, since they usually don't require any special administrative or financial arrangements, unlike with centralised facilities. P2P systems recommend an approach to aggregate and make use of the incredible computation and storage resources that otherwise just sit idle on computers across the internet when they are unused. P2P systems are widely used for file-sharing. The fundamental idea of file sharing is to utilise the idle disk space for storage and the existing network bandwidth for search and download [1]. A major benefit of P2P file sharing is that these systems are fully scalable—each additional user brings extra capacity to the system. In a P2P system, participating nodes mark at least part of their resources as 'shared', allowing other contributing peers to access these resources. Thus, if node A publishes something and node B downloads it, then when node C asks for the same information, it can access it from either node A or node B. As a result, as new users access a particular file, the system's capability to provide that file increases [2].

There are mainly three different architectures for P2P systems: *centralized, decentralized structured* and *decentralized unstructured*. In the centralized model, such as Napster [3], central index servers are used to maintain a directory of shared files stored on peers with the intention that a peer can search for the location of a desired content from an *index server*. On the other hand, this design makes a single point failure and its centralized nature of the service creates systems susceptible to denial of service attacks.

Decentralized P2P systems have the advantages of eliminating dependence on central servers and providing freedom for participating users to swap information and services directly between each other. In decentralized structured models, such as Chord [4], Pastry [5], and CAN [6], the shared data placement and topology characteristics of the network are strongly controlled on the basis of distributed hash functions. In decentralized unstructured P2P systems, such as Gnutella [7] and KaZaA [8], there is neither a centralized index nor any strict control over the network topology or file placement. Unstructured P2P systems also called pure P2P systems are most frequently used in Internet. Nodes joining the network, following some loose rules, form the network. The resulting topology has certain properties, though the placement of objects is not based on any knowledge of the topology [9]. The decentralization makes available the opportunity to utilise unused bandwidth, storage and processing power at the periphery of the network. It diminishes the cost of system ownership and maintenance and perks up the scalability.

P2P systems continue to grow according to recent measurement studies [10, 11, 12, 13]. These studies show that the bandwidth consumed by the P2P file-sharing applications has exceeded that of the WWW applications. Most of the P2P file-sharing systems that rely on voluntary donations from individual participants potentially face the problem of free-riding. Free-riders utilise the resources of a system while contributing not anything to the system. Users who attempt to benefit from the resources of others without offering their own resources in exchange are termed free-riders [14]. The free-riders are selfish nodes which only utilize other peers' resources providing none or limited contributions in return, have greatly jeopardized the fairness attribute of P2P networks. Pure peer-to-peer systems are completely decentralized and resources are shared directly between participating peers, the consequences of free riding are very terrible. In 2000, a measurement study of the Gnutella file-sharing network [15] found that approximately 70% of peers provide no files and that the top 1% of the peers provide approximately 37% of the total files shared. Similar patterns have been observed in subsequent studies of Napster and Gnutella networks [16]. A different study presented in [17] found free-riders have increased to 85% of all Gnutella users.

Two different issues affecting the P2P system with free-riders are discussed in [18]. Due to free riding, the number of objects in the P2P system shrinks or grows very slowly. As number of popular files is decreased the users' interest in the P2P system is drastically reduced so that the users eventually pull out of the system. When users who share popular files jump out of the system, the system turns out to be inferior in terms of the quantity of objects shared. This ultimately led to the fall down of the entire P2P system. The second issue is that the majority of query requests from different nodes are directed towards a few peers holding popular files. Hence, majority of the downloading requests are directed towards those overloaded peers causing network congestion. As the systems' main routine activities are seriously affected, such peers leave the P2P system gradually. Not only do free-riders deteriorate the quality of service for other peers, but they also intimidate the survival of the entire system. These issues point to the need for a rigorous mechanism for effectively managing the free riders in decentralised unstructured P2P systems.

This paper proposes a Q-learning based approach for handling free riders in decentralised unstructured P2P networks. Thus, the proposed scheme- Q-feed eliminates the probabilistic or heuristic approach by a learning based scheme. It utilises the past performance of nodes in the network for classifying a node as free rider or not. The proposed scheme mainly positions a neighbour of a peer in three states based on its performance. The idea is that the classification of neighbours based on their status restricts the amount of service being received from the neighbours so that the free riding behaviour can be greatly reduced. It also diminishes the services being provided by a peer to its neighbour. At any time a neighbour can move to higher status if it has shown improved performance. A node is punished for its free riding behaviour. The proposed autonomous replication scheme called Q-replication aims to increase the availability and fault tolerance in unstructured P2P network. The Q-replication autonomously replicates the popular objects to well-performing nodes according to their past performance.

In this paper, the terms 'peer' and 'node' are used in an interchangeable manner. The P2P system model comprises nodes and files (objects). There are a few neighbouring nodes ($n_1, n_2, n_3…n_n$) associated with a peer 'p'. The peer classifies a neighbour as normal, suspended or dormant, based on the service it provides or receives. The term 'file' stands for any general content in a node or peer. A file can have more than one replica in the system. The number of neighbours connected (links) to a node is called its degree. The topology of P2P networks is modeled as a network with an undirected graph G whose nodes represent hosts and edges represent internet connections between those hosts. Nodes are usually very dynamic, where some can join and leave the network in the order of seconds whereas other nodes stay for an unlimited period. When a user requests a file, a search for the file is initiated and other nodes in the network need to be queried if the file is not available locally. A query is composed of one or more required words.

The remainder of this paper is organized as follows. Section II reviews the related work for controlling free riders. The proposed free riding solution is discussed in section III. An overview of the Q-replication technique is given in section IV. The major steps of Q-replication are described as an algorithm in section V. Section VI illustrates Q-replication by means of an example. Section VII describes the experimental setup for conducting simulations. Section VIII discusses the results of experiments conducted. Finally, section IX concludes the paper.

## II. Related Work

Though cooperation is a key to many P2P networks' survival and success, understanding it is difficult without efficient methods. To address this requirement, researchers have proposed several approaches to make P2P networks "contribution-aware" and thus combat free riding [19]. There are a few schemes cited in the literature for managing the free-rider problem based on incentives. Also there has been much research in P2P networks using a social structure to improve cooperation by providing good incentives.

A technique for managing the free-riders discussed in [20] considers a P2P network as a social structure where each peer behaves as a person in a society, making judgmental decisions about other members in the society. The technique utilises the transfer of credit between peers to decrease the path length in queries. A selection strategy is proposed that involves different aspects of peer interactions in P2P networks. The credit transfer mechanism assists to deject mischievous peers by confiscating credits that they have with good peers and moving them to more cooperative ones. Peers that do not cooperate are eventually isolated from those that do cooperate. Further to that, the information about a credit transfer can be authenticated and verified.

In [21], a social network is used to model a P2P system. The P2P network is modelled using a directed graph, where the nodes are peers and the edges are connections between peers. There is a friendship between two peers which is represented by the directed edges in the graph. Each edge has a credit and a payment weight assigned to it, where the credit from one node to another is the payment from the other node to itself. The information about the data transferred between peers is used to describe the strength of the friendship. A balance of friendship is used in a decision function to determine routing paths. The algorithm is robust in a large population and relative high turnover rate. One potential problem for this incentive mechanism lies on the repeated transfer of data along the transaction path, as it may impose extra burden on system performance and network bandwidth if the mean length of path is long.

A completely decentralised system that allows efficient sharing of bandwidth in cooperative content sharing network is discussed in [22]. The system is called 'Scrivener'. It only requires nodes to track their neighbour's behaviour. It uses a greedy randomized routing algorithm to find a credit path, allowing a node to leverage credit it has with its overlay neighbours to obtain content from an unrelated node that holds the desired content. Scrivener is scalable and prevents freeloaders from exploiting obedient nodes.

In the simple modeling framework discussed in [23], a user in the network is a rational agent with a private and intrinsic characteristic called her type. Users make a decision whether to contribute to a system based on the association linking the cost of contribution and her type. This single parameter mirrors the readiness of the user to contribute its resources. If there are too few providers, then the deciding peer will be less eager to participate on account of the increased load on itself. The rate of contributing, to a peer, is the inverse of the total percentage of providers in the system.

The concept of utility function based scheme to control free riding in a P2P file sharing system is introduced in [18]. This utility-based scheme creates incentives to motivate users to share interesting files. This method doesn't allow peers to download files if their utility value is lower than the size of the requested file. Three important utility factors for building fair incentives in the file-sharing context are identified: the total number of files shared, the total size of data shared, and the popularity of the data shared. The authors claim that the proposed scheme can increase the lifetime of the system by 10 times. However, the technique completely relies on the precise information on peers provided by the peers themselves. Hence, malevolent client programs can easily cheat the network and spoil the anti-free riding measures.

A scheme that keeps track of the resource utilisation and resource contribution of each participating peer is proposed in [24]. In general the position of each participant in the system is denoted by a single scalar value, called their 'Karma'. Each node has an associated bank-set that keeps track of the node's Karma balance, which is an indicator of its standing within the peer community. The bank-set permits a resource consuming operation by the node only if the node has enough Karma in its account to allow the action. Karma doesn't require any centralized functionality or trust. Every time a peer's Karma changes, a predefined number of these peers should be reachable. Consequently, the ID of the peers should be known and not be transient. However, unstructured P2P networks do not maintain permanent and trustworthy identification techniques [25].

In a scheme proposed in [26], each node is associated with two parameters: money and reputation. Peers exchange money for service and increase their reputations while doing so. There is a central authority that settles disputes between peers when one believes it overpaid or did not receive enough service. The central authority is a set of randomly chosen nodes in the network. Similar to other schemes, they classify peers into three different types: honest, selfish, and malicious. As the solution relies on a centralised authority, its malicious behaviour will cheat other nodes connected to it. Also the dynamic nature of nodes in unstructured networks makes the technique an unreliable one.

There are different categories of methods for controlling free-riders. A classification of free riding techniques is presented in [19]. The schemes are categorized as monetary-, reciprocity-, and reputation-based approaches. Monetary-based approaches charge peers for the services they receive. Because these services are still very low cost, such approaches are also called micro payment-based solutions. The technique proposed in [24] is an example for monetary based approach. The main disadvantage is that the proposed solutions require some centralized authority to monitor each peer's balance and transactions. This can cause scalability and single-point-of-failure problems. In reciprocity-based approaches, a peer monitors other peers' behaviours and evaluates their contribution levels. The well-known P2P application BitTorrent, implement a reciprocity-based approach by adjusting a peer's download speed according to its upload speed. Reciprocity-based approaches face several implementation issues such as fake services published by peers. Since peer itself provides contribution level information the credibility is in question. In reputation-based approaches peers with good reputations are offered better services. These approaches construct reputation information about a peer on the basis of feedback from other peers. Reputation-based approaches store and manage long-term peer histories. XRep [27] is an example of an autonomous reputation system. Reputation sharing is achieved in XRep through a distributed algorithm by which resource requestors can evaluate the consistency of a resource offered by a participant before beginning the download.

The distributed framework proposed in [25] primarily focuses on locating free-riders and taking actions against them. Each peer monitors its neighbours, decides if they are free-riders, and takes appropriate actions. The free-riders are classified as non-contributors, consumers and droppers. This method does not need any interminable identification of peers or security measures for providing a global reputation system. Each peer just stores information about the neighbours' messages which are routed through it and executes the same kind of mechanisms alone and does not depend on any other peer's cooperation. The various counter measures proposed in the paper do not suggest utilising any score value for a peer's usefulness to the P2P system. One solution proposed to limit the network traffic is modification of TTL value based on the type of free-rider. Even for large TTL values, a simple mechanism of TTL reduction is employed and the reduced TTL value is always fixed for different categories of free-riders. The second type of punishment is dropping of queries originated from neighbours identified as free-riders. Another limitation of this technique is that the increase in availability of objects in the peers due to downloading and replication are not considered. Our technique reduces the value TTL of the low performing nodes based on its current TTL value Also, the popular objects are replicated to well-performing nodes autonomously.

## III. Q-Feed

The proposed solution 'Q-Feed' is intended for reducing the free riding effect in unstructured P2P networks. It widely employs Q-learning concepts so that the accuracy in managing free-riders is improved significantly. Each node in the network maintains a few data structures called Q-tables. Q-tables contain a list of Ids of neighbouring nodes of a peer, and their corresponding Q-values. It also contains the present status of a neighbour as normal, suspended, or dormant state. For each possible action, the Q-learning agent maintains a Q-value that indicates the efficiency of a P2P node in the past. A node in the network thus lives in any one of the three states. The status of a peer indicates the strength of a peer as a good node or not. A good node in the network is a well-performing node in the sense that it hosts more number of popular objects and it whole heartedly participates in the resource discovery process to improve the network performance. It never creates unnecessary traffic. The availability of the node may be high so that it can greatly serve other nodes in the network. In short, a good peer is altruistic and possesses a positive approach for serving other nodes. Hence, a node which is in the 'normal state' is a trustable peer as long as its Q-value is within a certain level.

Table 1. Q-Table of a node

| Node-id | $N_1$ | $N_2$ | $N_3$ | $N_4$ | $N_5$ |
|---------|-------|-------|-------|-------|-------|
| Q-value | 150   | 80    | 35    | 245   | 70    |
| Status  | Normal | Suspended | Dormant | Normal | Marked dormant state |

The status of a neighbour, for the most part relies on the corresponding Q-value in a Q-table. A Q-table contains the ID of each neighbour along with its Q-value and status (Table 1). The Q-value of a node is modified using a few parameters, which are collected periodically. The parameters are collected based on various actions of nodes and outcome of these actions. A node in one status can move to the next lower state and vice-versa when certain criteria are met. A node showing deteriorating performance first moves to the suspended state and then goes to the dormant state. But when the performance is improved, the node can lift its status. A suspended node can move to the normal state or a node in dormant status can move to suspended status after satisfying a few conditions. However, a much low-grade show of a node directs it to the dormant status. Contradictory to this degradation, a node possessing dormant status can also shift to normal status based on its greater performance. Marked dormant state is a special condition to be met by a dormant node to accomplish either suspended state or normal state. A free-rider peer is either in the suspend state or dormant state. The status thus explains the extent at which the free riding behaviour dominates in the peer. The access of network resources for such nodes is limited. Hence, a tight control mechanism is employed for dormant nodes than a node in the suspended state. A Q-value greater than or equal to 100 positions a neighbour in the normal state. A value less than 100 and greater than or equal to 60 shifts a neighbour to suspended state. A peer moves to dormant state if its Q-value in the Q-table is lower than 60. An example of a Q-table for a node is shown in Table 1.

Before initialising the Q-value for a neighbour in the Q-table, the peer P collects from its neighbour N, the number of files present in the shared folder $f_{current}$, the upload bandwidth $U_{band}$ and download bandwidth $D_{band}$. Using these values, the initial Q-value is computed as
$$\left[\left(u_1 * \left(\frac{U_{band}}{D_{band}}\right) + u_2 * f_{current}\right) * 100\right],$$
such that $u_1 + u_2 = 1$ where $u_1$ and $u_2$ are two weights, and $u_2 > u_1$ to give higher value for the second component in the equation. Thus, based on the number of files shared and values of upload and download bandwidth, an initial Q-value is assigned for the neighbour in the Q-table of P. For higher upload, a larger initial Q-value is assigned. Larger upload bandwidth allows a node for the faster sharing of files hosted by it to other nodes in the network. At the same time, the presence of large number of files in the shared folder increases sharing of files a large extent. The initial value is computed when a P2P node joins the neighbour group of a peer. However, in this work, a minimum Q-value of 100 is assigned for all the neighbours possessing initial Q-value less than 100 to acquire normal status, otherwise the nodes may be directly shifted to other states based on initial Q-value. The assignment of initial value gives opportunity to a node to show its performance and thereby receives rewards for the actions.

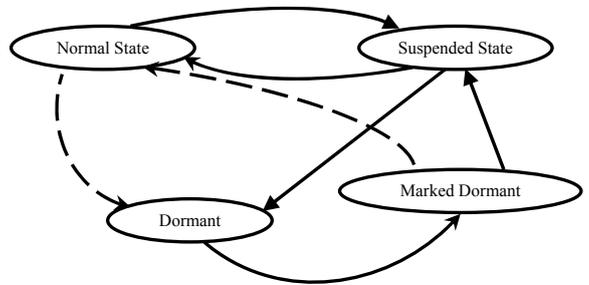

Figure 1: A neighbouring node of a peer in different states

The entire process of status transformation of a node in the P2P network is depicted in figure 1. The different techniques that are followed by each node in different states for managing free-riders are explained in the following parts of the paper.

*A. Normal State*

In the normal state, a peer doesn't enforce any restrictions for selecting a neighbour for forwarding the queries. A node in the P2P network is initially in the normal state. Several performance parameters are collected continually and based on these a node either stands in the normal state or moves to lower state. The chief deciding factor for changing the status is the Q-value of the neighbour in the Q-table of the peer. A neighbouring node in the normal state should always maintain its Q-value greater than a threshold value $L_{th}$. Otherwise; the node is degraded to lower status. For a particular period of time 't', a peer of a neighbour collects the values of parameters listed in Table 1 and computes the reward for all the actions in this period (Equation 1). The reward is used to modify the Q-value of a neighbour N of a peer P (Equation 2). The value of learning rate constant α is preset at 0.2. Such low value assigned to learning rate constant provides gradual increase of Q-value. The Q-value is modified using the computed reward and current Q-value. Higher the reward, an increase in Q-value of a neighbour in the Q-table will occur. The sum of all weights is equal to one such that $w_1 > w_2, w_3, w_4$. A large value for $w_1$ provides high priority for number of hits occurred in neighbour and number of results produced. The more number of hits near the peer decreases the number of hops to be travelled by a query.

$$\rho = \left[\left(w_1 * \left[\frac{PNHIT}{PN}\right] * AvgR + w_2 * \left[\frac{ONHIT}{ON}\right] + w_3 * \left[\frac{NA}{(NR-NA)}\right] + w_4 * \left[\frac{NGHIT}{NGQ}\right]\right) * 100\right] \ldots\ldots(1)$$

$$Q_{i,t+1} = Q_{i,t} + \alpha(\rho - Q_{i,t}) \ldots\ldots(2)$$

For all the neighbours, P maintains the values for all the variables listed in Table 2. The variable PN contains the number of requests generated by a peer P to a neighbour N for the specified period. P receives messages from other neighbours and sometimes these messages may be forwarded to N. The number of such messages is stored in a variable called ON. Similarly, the number of query hits for messages PN and ON queries are recorded. These are represented as PNHIT and ONHIT respectively. A node specialised in a particular area responds with more number of results for a query matching the area of specialisation. Hence, the number of results for queries submitted by P to N are monitored and at the end of the period, an average of number of results for all the queries submitted by P to N is computed (AvgR).

Table 2. Parameters collected from a neighbour in the normal state

| |
|---|
| PN - Number of query requests originated from a peer P to a neighbour N |
| ON - Number of query requests received from other neighbours and forwarded by P to neighbour N |
| PNHIT-Number of query hits from neighbour N for queries originated from peer P |
| ONHIT-Number of query hits from other neighbours for queries forwarded by P to neighbour N |
| NF-Number of files (objects) hosted in each neighbour |
| NR-Number of files replicated to a neighbour N by P |
| NA-Number of replicated files actually present at time t in neighbour N |
| AvgR-Average number of results per query for the requests sent to N for the period |
| NGQ – number of queries originated from neighbour N towards P |
| NGHIT – number of hits for queries originated from N through P (including hits at P) |

Because, Q-replication algorithm replicates copies of objects to nodes in the network, peer P may also receive and replicates objects. The number of files replicated during 't' by P to N is recorded (NR). The small portion of secondary storage is spared by a node for hosting shared files for the P2P network. All the nodes in the network can access these files by a resource discovery process. A free-rider node may intentionally remove the replicated files from its shared folder to save network resources. To verify this, P retrieves the number of replicated files present at N and the value is stored in NA. A neighbour also dispatches its own query messages (NGQ) to P and the number of hits for those queries are observed (NGHIT). Using the modified Q-value the status of neighbour as normal, suspended or dormant is decided. If the Q-value is less than a low threshold value, the neighbour node is shifted to *suspended state* and the extreme reduction of Q-value value moves the node to *dormant state*.

*B. Suspended state*

Assume a maximum TTL value is preset by the P2P system for all the nodes in the network. A neighbour N of a peer P is degraded to suspended state when the corresponding Q-value for the node in the Q-table of peer P reaches a value less than a threshold value, i.e. $Q_{i,t} < L_{th}$. For the queries received from suspended node, the peer P checks its shared folder for a query match and if the query is not satisfied at the peer, the current TTL value is decreased and the query is forwarded to a neighbour of the peer. The TTL value is modified as $TTL_x = round(\ln(TTL)), TTL_x \geq 0$, where $\ln(TTL)$ is the natural logarithm of current TTL value and $TTL_x$ is the modified TTL value which will be always less than current TTL value. Because the low TTL value causes reduction in network traffic, the messages associated with suspended nodes are significantly reduced.

The peer P records the number of queries originated from the suspended peer ($n_1$) for a certain period of time. If the value of $n_1$ is greater than a certain limit, Q-value for the suspended node in the Q-table of P is modified using reward computed. The reward calculation utilises number of files present in the free-rider, $f_{after}$ and the number of files present at the time of suspension, $f_{free}$. $f_{after}$ includes number of the old files as well as newly added files in the suspended node. The data is collected by sending a message to the suspended node. The reward is computed as $\delta = (w_1 * (f_{after} - f_{free})) * 100$ where $w_1$ is a weight and its value is between 0 and 1. The high value, which is assigned to w1, makes the reward high. This provides high priority to the increased availability. The Q-value of the suspended free-rider is modified as $Q_{i,t+1} = Q_{i,t} + \alpha(\delta - Q_{i,t})$, where α is the learning rate constant [28]. Due to this amendment, the Q-value is increased/decreased slowly for the free-rider based on number of new files being hosted in it. Thus, a node which is placed in the suspended state can gradually move to the normal state or lower state. The moment a neighbour moves to the suspended state, the peer P stops sending messages to it.

### C. Dormant state

Immediately, the Q-value in the Q-table for a neighbour reaches a threshold value $U_{th}$, i.e. $Q_{i,t} < U_{th}$, the neighbouring node is despoiled to the dormant state. Hence, the peer for which the node is a neighbour allows the free-rider very limited access to the network resources. The value of $U_{th}$ is always far less value than $L_{th}$. Any queries originated from the free-rider neighbour and the queries routed by it are processed by the peer and the queries are not routed further. If the required object is found, the querying node is informed. Otherwise, the queries originated/routed from the free-rider neighbour are dropped. However, the dormant nodes are low performing nodes because of low object availability and these nodes consume undue network resources by sending messages to other nodes for resource retrieval. The heavy flow of incoming query messages increases the load of a peer and thus the performance of the node may be seriously damaged. In this situation, the load on a peer should be effectively managed. The CPU usage of the peer is measured and if it reaches a maximum CPU-load, the queries arriving from dormant nodes are discarded. The maximum CPU-load threshold value is preset by the peer itself.

The major blow to a free-rider is that the moment a node is declared as dormant, the peer for which dormant node is a neighbour stops sending further query requests to it. Hence, once the node demeans to the dormant state, the peer of the node does not further utilise the resource discovery feedback for manipulating the Q-value of the dormant node in its Q-table. A node in the suspended state is more trustable than a node in the dormant state. Hence, it can simply shift to the normal status by just increasing the availability. But, the dormant node shows very low performance in file sharing because of its low performance shown earlier and so rigid policies should be followed for managing its status change. A node in the dormant state can move to suspended state or normal state if the present Q-value is increased to appropriate levels. This is done in two stages: the first stage employs a polling process and the second stage utilises both the object availability and the results of resource discovery.

In the first stage, the peer of a dormant neighbour does polling among other neighbours which are in normal status. The polling is done after a period of time preset by the system and it is a continuous process. Only well-performing nodes with higher Q-values are merely considered for the polling process. For polling, the average of Q-values of neighbours other than free-riders listed in the Q-table is computed (AvgQ). The neighbouring nodes whose Q-values greater than or equal to AvgQ are selected for conducting the poll. The peer dispatches messages to the chosen neighbours to gather the corresponding Q-value of the dormant node in their Q-tables. If the dormant node ID exists in the Q-table of a chosen node, the Q-value of the node is dispatched to the peer. A peer discards the message if id of the dormant node doesn't exist. The peer computes the average of the Q-values (AvgD) returned. If $AvgD \geq L_{th}$ dormant node is *marked* as a suitable candidate for moving to the suspended state. In this scenario a dormant node is said to be in *marked dormant state*. The present Q-value of the marked dormant node in the peer's Q-table is modified with the value of AvgD, i.e. $(Q_{i,t+1} \leftarrow AvgD)$. Even though, the Q-value after modification is greater than $L_{th}$ or $U_{th}$, the node remains live in the dormant state until certain conditions are met. This is to ensure that the node hosts some useful objects. If none of the selected node has the particular dormant node as neighbour, the Q-value of the node is not altered. The presence of higher Q-values for the node on other neighbours indicates the good service being it may provide to other peers. At the same time, nodes may deliberately collaborate to augment their Q-values. Thus, the result of polling may provide a Q-value which is ambiguous. To generate a feasible solution to overcome this limitation, the increase in availability of the objects and the quality of the objects are taken into account. The quality of the objects is to be considered as the node may attempt to increase the availability by hosting very low popular objects in its shared folder causing very low success rate. The high success rate for the queries inputted through suitable resource discovery mechanism reflects the quality of objects.

The second stage involves a two step process for the marked dormant neighbour for promotion to higher states. The first step is connected to the increased availability of shared objects. The second step is mainly intended for checking the usefulness of objects hosted in the dormant node. The marked dormant node can thus budge to the suspended state based on *increased availability of objects, and number of query hits*. The Q-replication relies on the node parameters such as bandwidth, degree of the node and free storage. There is a possibility that other nodes in the network running the autonomous Q-replication algorithm may replicate objects to the dormant node. The dormant node may also download objects from other nodes in the network after successful search operations. Thus, the availability of the objects may be reasonably improved. Hence, like in suspended state, the peer right now collects the number of shared objects available ($F_a$) in the shared folder of the dormant node. The number of objects exists in the node ($F_d$) at the time of degradation to dormant state is recorded earlier. The difference ($D_{da}$) between $F_d$ and $F_a$ is computed. The number of objects downloaded between the current time and the time at which the node is declared as dormant may be very low or nil in the shared folder of the node if the node exhibits free riding status for which the node is a neighbour for other peers. As mentioned earlier, the incoming queries from dormant nodes are not forwarded favourably to other nodes by the peer. If $D_{da} \geq [F_a + \ln(F_a)]$, the dormant node begins to receive limited number of queries from the peer for which the dormant node is a neighbour. The second component $\ln(F_a)$ ensures that nodes which are hosting additional objects other than those present at the time of degradation to dormant state are also chosen for the second stage of processing.

In the second step, not all the queries that are originated from P or forwarded by P towards the dormant node are processed by P. The Q-value of the marked dormant node is moderately higher due to modification with AvgD. For testing the quality of objects hosted in the marked dormant node, a small number of queries are routed periodically ($n^{th}$ query 20th, 40th, 60th...) by the peer. All the query hits happened in the dormant node are recorded. Hit ratio for number of queries is computed as a ratio between the number of hits and number of queries being processed. If the hit ratio ($H_r$) is higher than a threshold hit rate ($H_{th}$), the marked dormant node moves to the suspended state, otherwise, the process is repeated until the required hit rate is achieved. Thus, through multifaceted actions, a dormant free-rider modifies its status.

## IV. Q-Replication

Replication addresses one aspect of the free riding problem [18]. The objective of a replication technique is to improve availability and enhance system performance. Since a small number of nodes host popular objects, these nodes may receive large quantity of query messages from other nodes. A solution to increase the availability of objects in the network is replication. It makes use of the disk space and the network bandwidth resources of the node downloading a particular file. Thus, more number of nodes satisfy downloading requests from several nodes. This helps to reduce network congestion and CPU overloading problems that the small number of peers experience.

P2P-based replication strategy is the topic of various active research projects and most of these projects emphasize replication in structured P2P systems. Only a few replication techniques are cited in the literature for decentralised unstructured P2P networks. Merely a few replication techniques for unstructured P2P networks are cited in the literature. Most of the replication schemes are linked to the search techniques being employed, i.e. objects cannot be replicated to further than the nodes on the search path. In all cases, the behaviour of nodes is not taken into account while replicating the objects. Hence, to increase the availability and to replicate the objects autonomously, a novel replication technique is proposed in this work. Our scheme, known as "Q-replication" employs Q-learning for the autonomous replication of objects [28, 29, 30]. It is autonomous because the decision to replicate an object to appropriate sites is taken autonomously by a node based on the past performance of peers in replicating objects. Thus in spite of constant changes to the connection, objects are highly available. Like Q-Feed, the proposed replication technique maintains a Q-table which contains peer-IDs and corresponding Q-values of each peer. A Q-value represents how a peer has contributed to the replication activities in the past. As part of replication, a node receives a reinforcement signal from the target node which is intended for hosting the replica. The signal is translated into a reward. Parameters such as bandwidth, degree of the node, and storage cost are utilised. The Q-values are updated appropriately. The Q-replication process selects the target objects for replication based on their popularity. The popularity is computed in a unique way. After replicating an object, the Q-values corresponding to the nodes in the Q-table are updated using the various parameter values returned. A node which goes down frequently and maintains low values for bandwidth, degree, and available storage may produce small value for reward. These nodes show less performance in the replication process. Hence well-performing nodes receive high Q-values and the Q-values of nodes with low performance are reduced further. A shared object is replaced from a node to accommodate a new object by utilising the popularity and the time at which the object was inserted into the shared directory of a node. The Q-replication, thus, distributes the popular objects to well-performing nodes in the network for improving the availability of objects and thereby contributes to the improvement of success rate and fault tolerance without depending on search paths.

The Q-replication scheme is distributed, and employed without the coordination of centralized servers. It considers the replica selection problem (which data to replicate) and the replica placement problem (where to place them), and provides simple solutions to them. The replica selection problem deals with a suitable criterion for selecting an object from the shared storage space of a node for replication. The replica placement problem addresses the process of choosing an appropriate node for hosting a replica.

### A. Selection of Objects for Replication

The objects are chosen for replication according to their popularity. The frequently accessed objects from the shared storage space of a node are treated as popular objects. These files are ranked according to their popularity. The details are stored in a table, which contains the object name along with its rank and the status of replication. The value of the rank represents the popularity of the object. The high value rank denotes a most popular item. The status field facilitates to identify already replicated files in a node. The system regularly (e.g. for every 50 requests received by a node) updates the popularities of all the objects in the nodes based on incoming queries for that period. The popularity update process $P_f(t+1)$ at a time $t$ relies on the number of requests received for the object $R_q(t)$ and the total number of requests received by the node $N_q(t)$ up to that period after the previous update, and the present value of popularity $P_f(t)$. The popularity is modified as

$$P_f(t+1) = P_f(t) + \eta\left[\left(\frac{R_q(t)}{N_q(t)}\right)*100\right]$$

where the value of $P_f \geq 0$ and the value of constant $\eta$ is in between zero and one. The values thus modified are written into a table (*popularity table*) after removing the existing values. The update equation shows that the update process also utilises the existing popularity value for modification. The initial value of $P_f$ for an object is always zero. If the number of queries received for the time period is nix, the popularity of the object is not altered. Otherwise, the popularity of the file increases with the number of queries being received. For every $\delta$ period, the system identifies the possible candidates for replication. This is done by comparing the popularity of a file $P_f(\delta)$ with a threshold popularity value $P_{th}$. When the popularity of a file at $\delta$ becomes greater than or equal to the threshold value, i.e. $(P_f(\delta) \geq P_{th})$, the process of selecting the target nodes for hosting the replica is initiated.

*B. Q-table creation and Initialisation*

The target nodes are selected from the Q-table. The members for the Q-table are assigned after a simple operation: a message (Hello message) is sent to nodes that come within a time-to-live (TTL) limit, which is the number of hops the message should be propagated; the responded nodes become members of Q-table with some initial Q-value. The message forwarding follows a k-random walk [9] procedure. Initially K messages are generated and the messages are propagated through K number of neighbours selected randomly. Neighbouring nodes forward the message to one of their neighbours; from there to next hop. The message has a message-id. Nodes, which have already received a copy of the message, keep the message-id and address of the neighbouring node to which the message was forwarded. Hence, when a node receives the same message another time it will not be forwarded to a node that has received the message previously, but selects a different peer from the neighbour list. The response messages from the peers consist of equivalent values for their current bandwidth $b_w$, and available storage $s_{avbl}$. Using these values, Q-tables are initialised. The P2P system assigns minimum values for node attributes such as bandwidth $b_{min}$ and storage $S_{min}$, which are used for Q-value computation. The Q-values each node in the table is initialised as $Q_r = \left(\frac{b_w}{b_{min}} + \frac{s_{avbl}}{s_{min}}\right) * 100$. In order to eliminate the random or probabilistic assignment, Q-values are thus initialised with important node attributes bandwidth and storage.

*C. Selection of nodes and replication*

A good peer, which can host a replica, should have a high-speed connection, minimum available storage, link with more number of nodes and it should stay online for a long period. From the possible set of host candidates listed in the Q-table, the best ones according to the bandwidth, available storage, and number of links (degree) are chosen. The objects are copied into nodes, which do not already host the same replica of the target file. Hence, the overwriting of the same file in a node is avoided and at the same time, the process saves bandwidth consumption due to redundant file transfer. In order to choose the possible candidates for hosting the replica, the mean of Q-values listed in the Q-table is computed. Nodes with Q-values greater than or equal to the mean (AvgQ) are selected and a message is sent to each selected node to verify whether a copy of the object exists in their shared folder. *Replication List* of a node is a table that contains a list of object names reserved by other nodes during the object checking process. This evades other nodes to replicate the same object to a node as the same node may be chosen by another node as a target node for replication. If the node is not up, a copy of the object is present, or the object's name appears in the replication list, the node is left out from replicating the chosen file. All other nodes, whose Q-values greater than or equal to AvgQ are selected as target nodes for hosting replicas.

*D. Reward computation*

The nodes, which received a copy of the file, send the values for degree $(d_d)$, bandwidth $(b_w)$ and available storage $(s_{avbl})$, after storing the replica to the node that initiated the replication process. This is the reinforcement signal to the replication system. Based on the reinforcement signal, the reward $(\rho_i)$ is computed for each node in the Q-table as

$$\rho_i = \left[\left(\left[\frac{d_d}{d_{min} * w_1}\right] + \left[\frac{b_w}{b_{min} * w_2}\right] + \left[\frac{s_{avbl}}{s_{min} * w_3}\right]\right) * 100\right]$$

, where $w_1 + w_2 + w_3 = 1$. As the bandwidth is a very important network resource, priority is given for it while computing the reward, hence $w_2 < w_1, w_3$. Therefore, the nodes with large bandwidth highly influence the reward. Moreover sufficient storage space should be available in a node for hosting more and more replicas of different objects. In a P2P network, a few nodes have a large number of degrees while most of other nodes have only a small number of degrees. Peers with a large number of degrees make many replicas as peers with a small number of degrees [31]. In addition, replicas on large degree peers are used frequently as those on peers with small degrees. In our strategy, the system assigns a common minimum degree threshold $(d_{min})$ value to be used for replication to all nodes. In terms of *degree*, the contribution of high degree nodes to the reward is high as compared to low degree nodes. At the same time, nodes with only high bandwidth and storage can also participate in the replication process. All these factors ensure the availability of objects within short hop distances.

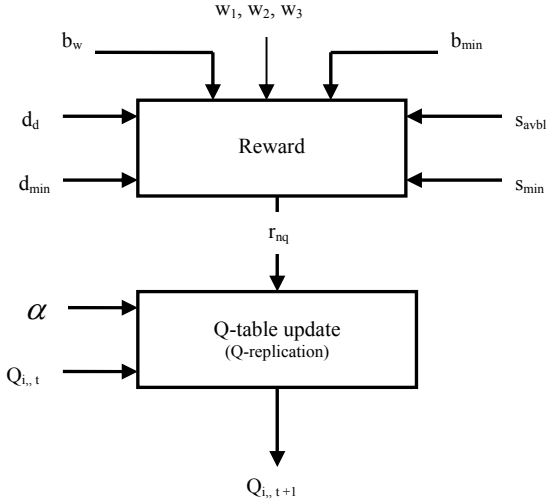

Figure 2. Q-table update (Q-replication)

## V. Q-Replication Algorithm

Here we present the entire algorithm for Q-replication for implementation. The Q-replication algorithm chooses the popular objects for replication whenever the popularity of the objects reaches certain threshold. The peers, which are n hops away, are included in a Q-table along with their past performance represented as Q-values. The target peers for hosting the replicas are selected from the peers listed in the Q-table according to the Q-values of nodes. Peers with Q-value greater than or equal to the average Q-value are chosen as target nodes. The shared directory of each chosen peer is searched for the presence of the selected object, and based on the result, the object is replicated. The replication process returns reinforcement signals comprising available storage, bandwidth and degree of the node. The reinforcement signal is converted into a reward. The Q-values of nodes in the Q-table are updated using the computed reward values. Finally a list of nodes which have received a copy of the replica is displayed. Q-replication is a greedy algorithm with a worst-case time complexity of $O(n)$, where 'n' is the number of nodes in the network.

### *E. Q-table Update*

The reward values are utilised to modify the Q-values (figure 2). The update process increase, or decrease the Q-values of peers that are being participated in the replication process. The nodes, which have not participated, do not modify the present Q-values. The nodes with high Q-values are treated as good peers. The Q-values of nodes, which have created a replica, are updated as $Q_{i,t+1} \leftarrow Q_{i,t} + \alpha(\rho_i - Q_{i,t})$, where $\alpha$ is the learning rate (value of $\alpha$ between zero and one), and $Q_{i,t}$ is the present Q-value. If the reward of replication is high, the Q-value is incremented and it relies on bandwidth, available storage and degree of the node. The current Q-values are retained for the nodes comprising a copy of the object i.e. $Q_{i,t+1} \leftarrow Q_{i,t}$. Nodes that are not up are punished heavily with zero reward, $\rho_i = 0$ and the Q-values are updated as $Q_{i,t+1} \leftarrow Q_{i,t}(1-\alpha)$. Assigning high value to the learning rate constant yields a large increase in Q-values of nodes that have placed a replica to their respective directories.

### *F. Object Replacement*

Some replicas should be deleted to make space for new replicas if adequate storage space is not available in a node. Our replication scheme removes the objects according to their popularity and age. The age attribute represents the time at which the object was inserted into the directory. If the object is recently added to the shared directory, it may have low popularity value and small value for age. Hence, objects with low popularity values and large values for age are removed for housing new objects.

### *Algorithm:*

Input: Object 'f' for replication; Q-table.
Output: a list of nodes, which have received a copy of the replica.

1. Select an object *f* for replication based on its popularity.
2. Compute the average of Q-values corresponding to Q-table of node x - AvgQ.
3. For each entry Ti in the Q-table, select nodes with Q-values >= AvgQ.
4. For each selected node in step 3, check for the presence of the object.
5. if *f* exists in the searched nodes or node is not up or the object's name is in the *Replication List*, leave out nodes from replication process
   else
       5.1 Select the remaining nodes with Q-values >= AvgQ for replication.
       5.2 Insert the object's name in the *Replication List* of *the* selected nodes.
6. For each chosen node N:
       6.1 Replicate the object from source node to target node.
       6.2 Remove the object entry from the *REPLICATION LIST* of the node, which has received the replica.
       6.3 Wait for reinforcement signal.
       6.4 Receive reinforcement signals—available free storage after storing the file, bandwidth and degree of the node.
       6.5 Compute the reward using the parameter values:
   $$\rho_i = \left[\left(\left[\frac{d_d}{d_{min}*w_1}\right] + \left[\frac{b_w}{b_{min}*w_2}\right] + \left[\frac{s_{avbl}}{s_{min}*w_3}\right]\right)\right]*100$$
       6.6 Update the Q-value of node, which has received a replica as: $Q_{i,t+1} \leftarrow Q_{i,t} + \alpha(\rho_i - Q_{i,t})$
       6.7 Nodes with a copy of the object, which are excluded in step 5, do not alter their Q-value, $Q_{i,t+1} \leftarrow Q_{i,t}$
       6.8 Nodes that are not up (step 5), accept zero reward and update Q-values as $Q_{i,t+1} \leftarrow Q_{i,t}(1-\alpha)$.

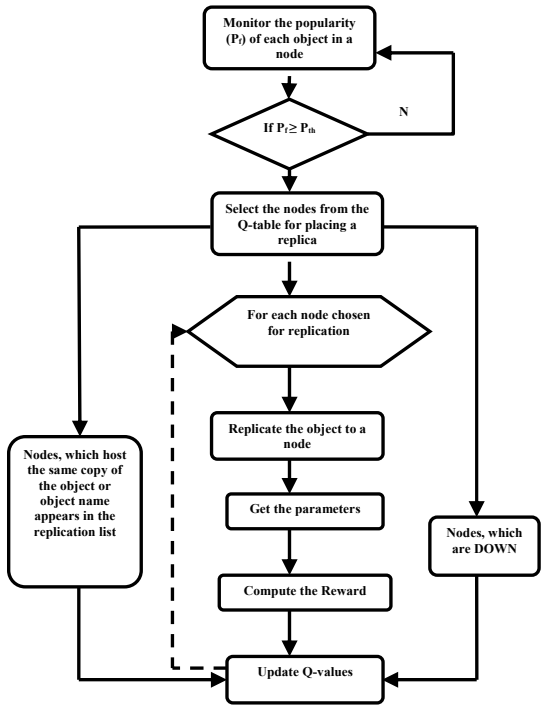

Figure 3. Set of actions taken by a node for replicating an object

## VI. Applying our Approach (Q-Replication)

This section illustrates the autonomous replication technique with a simple example. The results shown in the example is only for illustration purpose only and it does not reflect the situation in a real network. A simple P2P network in a graphic representation is shown in Figure 4. The neighbours of node A are nodes B, C and D. A 'HELLO' message is sent with two walkers to nodes within three hops away. The walker messages are fired though the neighbours B and D. From the second hop onwards, only one neighbour of each node is inserted into the replication Q-table. As a result, the Q-table (Table 3) of A contains neighbouring nodes as well as the nodes H, N, E, and F. Based on the attribute values; the Q-table is initialised. The Q-values are modified after each replication action.

Consider a scenario in the network after a few replication operations. There is an object '$f_1$' available for replication at node 'A'. Assume the object is present in none of the nodes listed in the Q-Table, and all the nodes are 'up'. The average of Q-values is found (*AvgQ=364*) and the target nodes are selected from the Q-Table (Table 4). The nodes B, D and N have the Q-value greater than equal to AvgQ. However, the node 'D' is not up; hence, it receives a negative reinforcement. The values of various attributes are collected and the rewards are computed. The value of $\alpha$ is preset as 0.2. The object is replicated to the nodes B and N (Figure 5). The Q-values are updated according to the update policy. The modified values in the Q-table of node A is shown in Table 4. The update operations increase the Q-value of both the nodes B and N. As the status of node D is 'down', its Q-value is diminished. All other nodes, which have not participated in the replication, keep the values as such.

Table 3. A Replication Q-Table

| Neighbours and nodes 3-hops away | B | C | D | H | N | E | F |
|---|---|---|---|---|---|---|---|
| Initial Q-values | 482 | 568 | 345 | 234 | 132 | 324 | 185 |

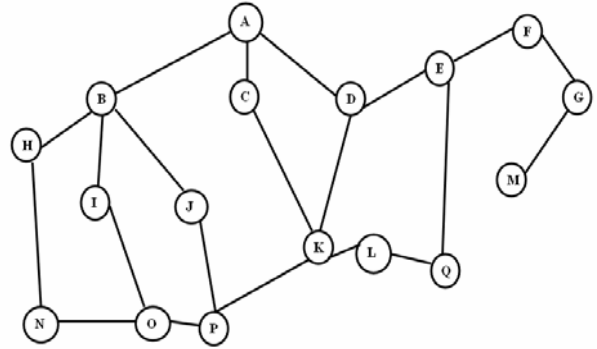

Figure 4. An Unstructured P2P Network

Table 4. Status of Q-table of node 'A' before and after replication

| Neighbours and nodes within 3-hops away | | B | C | D | H | N | E | F |
|---|---|---|---|---|---|---|---|---|
| Q-values after a few operations | | 682 | 122 | 441 | 336 | 466 | 324 | 175 |
| Parameters | Storage, $s_{avbl}$ $s_{min}= 40$ | 75 | 45 | 60 | 63 | 65 | 69 | 47 |
| | Bandwidth, $b_w$ $b_{min}= 64$ | 98 | 42 | 71 | 77 | 92 | 73 | 32 |
| | Degree, $d_d$ $d_{min}= 3$ | 3 | 2 | 3 | 1 | 2 | 3 | 2 |
| Node Status | | up | up | down | up | up | up | up |
| Reward ($w_1$=0.4, $w_2$=0.2, $w_3$=0.4) | | 1485 | - | 0 | - | 1292 | - | - |
| Updated Q-values | | 843 | 122 | 353 | 336 | 631 | 324 | 175 |

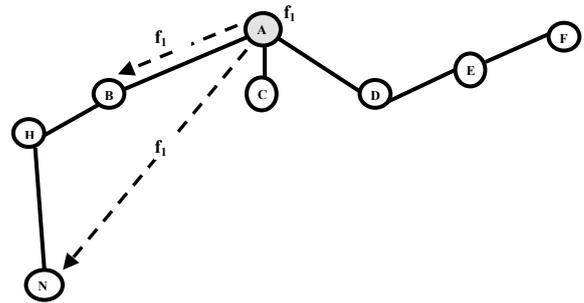

Figure 5. Replication of object 'f1' to nodes B and N

## VII. Experimental Setup

For improving the resource discovery process, popular objects are distributed to different nodes in the network.

Hence, the efficiency of Q-feed relies significantly on Q-replication as the status change of a node is very much linked to the availability of popular objects in well-performing nodes. The proposed algorithms are simulated using random graphs that have 10000 nodes. The nodes can join the network and establish random connections to existing nodes. There are multiple copies of 3000 objects randomly distributed to all the nodes in the network. Since the files are assigned randomly, the number of files in each node is different. The average degree of a node in the network is 3.5. We assume that all categories of nodes after the configuration of the network are formed due to several conditions preset in the proposed algorithms. The popular objects are replicated to different sites using autonomous replication algorithm. The bandwidth and space for shared storage are assigned to each node randomly from a list containing possible values. The quantities of objects maintained in the system are sufficient to analyse the performance since autonomous replication scheme effectively propagates the objects to various sites. The objects are word, PDF and text files available as course materials on various subjects such as computer science, electronics, physics, mechanics, electrical etc. Thirty thousand keywords are chosen from the course material files and these words are randomly selected as query keywords by all the nodes during searching. Two types of query search are employed: *file name based and keyword based*. In the file name based search only the objects names in the shared storage space of each node is searched. The objects containing the keywords are looked up in the keyword-based search. In the simulation scenario, all the queries contain keywords alone. Because the files are randomly distributed the number of files in a node varies form another node. This number will change due to successful downloads and replication. In the beginning of simulation all nodes have the same status. Based on several parameters, a node moves to different status to reflect their contribution level in the network.

The 'k-random walk' [9] search technique is employed for resource discovery in the network. When a user enters a query, the shared folder of the node is searched for the presence of the required object. If object is not present, the query is further forwarded to the K number of walkers. From there onwards queries are forwarded only through one neighbour of a node in the search path. Neighbours which are in 'normal status' are the candidates for participating in the search process. However, queries are also propagated through marked dormant node according to the prescribed criteria. A counter records the number of queries being processed by a node. The search is terminated either result is found or time-to-live (TTL) is expired. The default TTL value is preset as six. The number of walkers for a query operation is fixed as six. If sufficient number of neighbours are not present, all the neighbours are selected without bothering about the maximum number. Each node generates 100 queries and one query is propagated every 20 seconds on average. However, each node enters the query generation phase in a randomly selected time slot. Hence, the flood of query message production is regulated. One by tenth of simulation time is utilised for each period of operation. Eighty percent of the nodes are up at the time of performing simulation. Fifty percent of 'Down' nodes selected randomly change their status to 'UP' after every 50,000 queries are propagated and, at the same time, the same amounts of UP nodes obtain the DOWN status.

The simulation tool has been developed using Java language. The tool runs in a Windows operating system environment. The software, which are used for developing the simulation software are NetBeans, J2SE Development Kit 5.0 and WampServer. NetBeans is a free, open-source Integrated Development Environment, which supports development of all Java application types. WampServer is an open source project and Windows web development environment. It allows creating various applications with Apache, PHP and the MySQL database. WampServer also includes PHPMyAdmin and SQLiteManager for managing databases. The simulations are conducted in systems with Intel xeon (Quad Core) processor,12 MB L2 Cache, 1333 MHz FSB, 4 GB, and 146GB SAS HDD(15K RPM).

## VIII. Results and Discussion

The experiments are conducted by running the simulation tool number of times. Data for different parameters are collected each time. We have not intentionally made any neighbour of a peer as free-rider in the network. A few performance metrics are employed to evaluate the performance of the proposed techniques.

*Number of simulation runs vs. % of nodes in different states:* All the nodes in the network are linked to one or more neighbouring nodes and the details about these nodes are maintained in the Q-table of a peer. One neighbour of a node may be a neighbour of several other nodes in the network. Hence, the same node may possess different status in Q-tables of other peers. The status of all the neighbours of a peer in different categories as normal, suspended and dormant are counted after each simulation run. This is repeated for all the peers in the network. The number of times a node exists in each status during a simulation run in the whole network is counted and the status with highest value is selected for further operations. Based on the values so computed, the percentage of nodes in different status is worked out for all runs. The data is collected from only the nodes which are up at the end of each simulation.

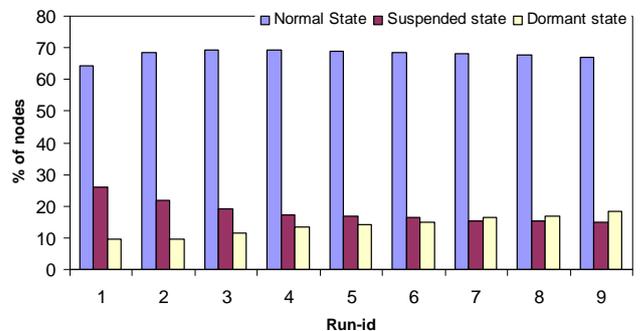

Figure 6. The nodes in different status

The situation for nine runs is plotted and shown in figure 6. The data in the previous run is kept as such for the next run. The free-riders are formed during simulation without any external intervention. So, based on the performance of neighbours of peers, the nodes with different status are created. Because of Q-learning, the network takes time to identify the free-riders in the network and each node in the suspended state or dormant state is allowed to improve their status according to the outcome of their actions. As shown in figure 6, after finishing five runs, the number of nodes possessing the normal status is going in a steady position. But the number of nodes in suspended state is slightly decreased and the number in the dormant state is increased slightly. Most of the low performing nodes are identified within a few simulation runs. This is due to the knowledge the nodes acquired through Q-leaning. The advantage is that a peer gradually identifies well-performing neighbours based on their Q-values and status for future operations. Further to that, Q-replication increases the availability of popular objects by replicating objects to nodes satisfying certain conditions stipulated by the replication algorithm.

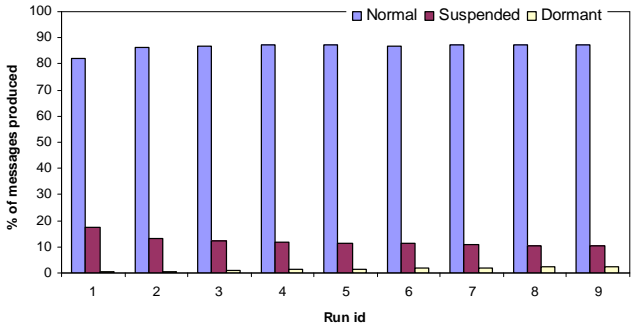

Figure 7. Messages produced for each run by nodes in different status

*Number of simulation runs vs. % of messages produced in each category of node:* Each peer records the number of messages received from each neighbour of the peer and the messages routed to each neighbour by the peer according to the neighbour status. After each run the total number of messages due to neighbours in different status for all the peers in the network is computed. Using the values for each category and the total number of messages for all categories of neighbours, the overall percentage of each category of messages is computed and plotted. This is shown in figure 7.

As shown in figure 7, the number of messages produced by nodes possessing normal state goes marginally in a steady state. But the messages from suspended nodes and dormant nodes are well managed from the initial run onwards. Hence, the positive messages originated from well-performing nodes dominate the network traffic so that messages generated unnecessarily from suspended and dormant nodes undergo tight control mechanisms available in Q-Feed.

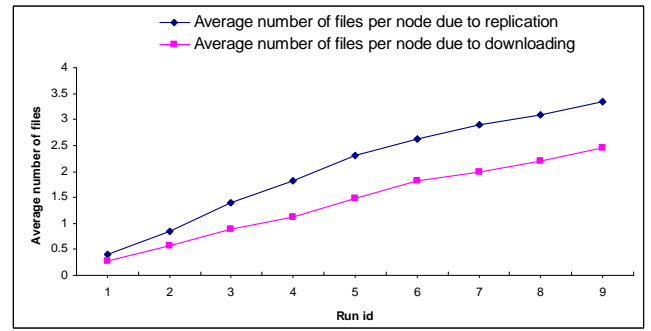

Figure 8. Number of files due to Q-Replication and Downloading

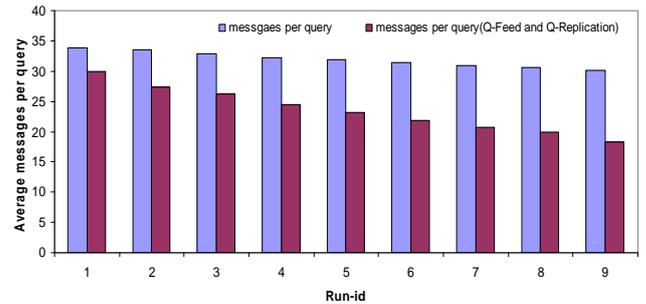

Figure 9. Average messages per query

*Average number of files due to Q-replication and downloads:* This experiment is conducted to compute the average number of files created as a result of autonomous replication scheme and downloading. The average number of files up to each run is computed and plotted as shown in figure 8. The number of files present in the nodes before replication is not counted. The number of files in each node increases with time. The proposed replication scheme, Q-replication is not related to the free-rider controlling scheme, Q-Feed. Hence, as popularity of a file increases, the number of copies of the file to be created also increases. Due to the spread of popular objects in the network, the query success rate also increases. Each query hit may also create another copy of the object in the query source by means of downloading from the node where query hit occurs. Hence, the availability of objects in the network also relies on query success rate. The Q-replication contributes more number of replicas of popular objects.

*Average messages per query:* The aim of this experiment is to find the average number of messages generated for each successful query for k-random walk with and without Q-replication and Q-Feed. The results are plotted as graph in figure 9. The messages per query are declining at a snail's pace for the case which doesn't employ the proposed techniques. The decrease in messages for the proposed solution relies on availability of popular objects due to Q-replication and the stringent measures of Q-feed for reducing the ill-effect of free-riders.

*Queries finished (k-random walk with path replication):* simulation experiments are conducted in a random network comprising 10000 nodes to compare the performance of path replication [9] and Q-replication on random K-walk search technique [9]. Path replication replicates an object along the path of a successful "walk". It doesn't cover nay other node in the network for hosting replicas. The number of walkers are limited to six. The results for queries finished in each simulation run are shown in figure 10. The success rate of random k-walk search with Q-replication is higher than the success rate produced for random k-walk with path replication technique. The influence of Q-replication in success rate improvement is very high as compared to path replication in each simulation interval. Each simulation run creates new replicas of popular objects in various nodes. Q-replication creates replicas of popular objects in more number of nodes and at the same time, path replication relies only the nodes on the search path in which the target nodes to host replicas are not selected based on their performance in the past.

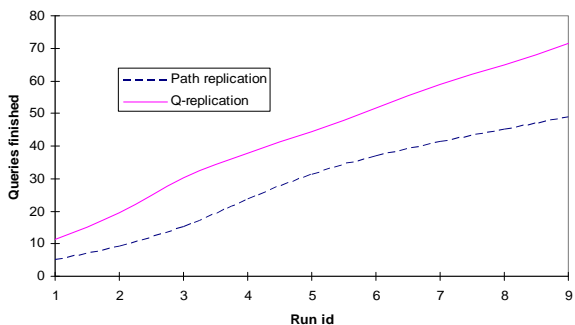

Figure 10. Percentage of queries finished

## IX. Conclusions

The proposed solutions for managing free-riders widely employ the Q-learning concepts. The Q-Feed algorithm is a greedy algorithm for controlling services among low performing neighbours of a peer. A neighbour may be in one among the four states at a time based on its services rendered to a peer. The nodes possessing low status are encouraged to increase their position. Simulation results show that Q-Feed effectively manages free-riders in the network. The average number of messages generated by for each successful query is significantly reduced. In addition, Q-replication productively replicates the popular objects among well-performing nodes and the algorithm assists the neighbours to prevail from free riding behaviour by hosting popular objects in the shared folder. Like monetary based techniques, no central monitoring agent is required for Q-feed. Q-feed encourages all free riders to attain higher status.

The proposed replication approach utilises the popularity of the objects and the objects are distributed with more copies to various sites based on site selection logic. The popularity is computed according to the queries received on a particular objects and the total number of queries received by the node for a certain period. The target nodes are selected not randomly nor probabilistically, but they are chosen based on their past performance. The replication scheme does not rely on nodes on the search path. Other nodes can also host the same replica of the object, provided that the sites satisfy certain criteria. The replacement of a file follows a different approach and it depends on its popularity and age.

## References

[1] Can Erten and Richard MacManus, March 29, 2007, P2P: Introduction and real world applications, *Read/Write Web*, available http://www.readwriteweb.com.
[2] Saurabh Tewari, Performance study of peer-to-peer file sharing, *Ph.D Thesis*, University of California, Los Angeles, 2007.
[3] Kim and L. Hoffman, "Napster and other Internet peer-to-peer applications," George Washington University, available: citeseer.ist.psu.edu/kim01pricing.html,2002.
[4] Stoica, Morris, R. Karger, D. Kaashoek, M and H. Balakrishnan, "Chord: A scalable peer-to-peer lookup service for internet applications," Proc. of SIGCOMM 2001.
[5] Rowston and P. Druschel, "Pastry: Scalable, distributed object location and routing for large-scale peer-to-peer systems," Proc. of IFIP / ACM Middleware, Heidelberg, Germany, 2001.
[6] S. Ratnasamy, P. Francis, M. Handley and R .Karp, "A scalable content-addressable network," In Proc. of SIGCOMM, 2001.
[7] The Gnutella Home Page: http://www.gnutella.wego.com.
[8] The KaZaA Home Page: Page. http://www.KaZaA.com.
[9] Lv. C, Cao. P, Cohen. E, Li. K and Shenker. S, "Search and replication in unstructured peer-to-peer networks," Proc. of 16th Int. Conf. Supercomputing, pp. 84-95, 2002.
[10] Mohamed Hefeeda, "Peer-to-Peer systems", School of Computing Science, Simon Fraser University, Surrey, Canada, 2004.
[11] K. Gummadi, R. Dunn, S. Saroiu, S. Gribble, H. Levy, and J. Zahorjan, "Measurement, modeling, and analysis of a peer-to-peer file-sharing workload," proc. of 19th ACM Symposium on Operating Systems Principles (SOSP'03), Bolton Landing, NY, USA, October 2003.
[12] S. Saroiu, K. Gummadi, R. Dunn, S. Gribble, and H. Levy, "An analysis of Internet content delivery systems," proc. of 5th Symposium on Operating Systems Design and Implementation (OSDI'02), Boston, MA, USA, December 2002.
[13] S. Sen and J. Wang, "Analyzing peer-to-peer traffic across large networks," IEEE/*ACM Transactions on Networking*, vol. 12, issue 2, pp. 219 – 232, April 2004.
[14] Michal Feldman and John Chuang, "Overcoming free-Riding behavior in peer-to-peer systems," ACM SIGecom Exchanges archive, Vol. 5, pp. 41– 50, Issue 4, July 2005.
[15] Eytan Adar, and Bernardo A. Huberman, Free riding on Gnutella, *First Monday*, Volume 5, Number 10 - 2 October 2000.
[16] Stefan Saroiu, P. Krishna Gummadi, and Steven D. Gribble: A measurement study of peer-to-peer file sharing systems, proc. of Multimedia Computing and Networking (MMCN) 2002, San Jose, CA, USA, January 2002.
[17] Hughes, D., Coulson, G., and Walkerdine, J., "Freeriding on Gnutella revisited: The bell tolls?" *IEEE Distributed Systems Online*, vol. 6, no. 6, pp. 1-18, June 2005.
[18] Lakshmish Ramaswamy and Ling Liu, "Free Riding: A new challenge to peer-to-peer file sharing systems," Peer-to-Peer Computing Track, Hawaii International Conference on System Sciences (HICSS-2003), January 2003.
[19] Murat Karakaya, Ibrahim Korpeoglu, and Özgür Ulusoy, "Free riding in peer-to-peer networks," *IEEE Internet Computing*, pp 92-98, 2009.

[20] V. Ponce, J. Wu, and X. Li, "An approach to combating free-riding in peer-to-peer networks," *International Journal of Parallel, Emergent and Distributed Systems*, 2009.

[21] W. Wang, R. Yuan, and L. Zhao, "Improving cooperation in peer-to-peer systems using social networks," Proc. of 20th IEEE Int'l Parallel & Distributed Processing Symposium, pp.446-453, 2006.

[22] Nandi, T. Ngan, P. Druschel, and D. Wallach. "Scrivener: providing incentives in cooperative content distribution systems," Proc. ACM/IFIP/USENIX Int'l Conference on Middleware, pp. 270-291, November 2005.

[23] M. Feldman, M. Ahamad, C. Papadimitriou, J. Chuang, and I. Stoica, "Free-riding and whitewashing in peer-to-peer systems," proc. of the SIGCOMM'04 Workshop, August 2004.

[24] Vivek Vishnumurthy, Sangeeth Chandrakumar, Emin Gun Sirer, "KARMA: a secure economic framework for P2P resource sharing," proc. of workshop on the Economics of Peer-to-Peer Systems, 2003.

[25] Murat Karakaya, _Ibrahim Ko¨rpeog˘lu, O¨ zgu¨ r Ulusoy, "Counteracting free riding in peer-to-peer networks" *Computer Networks*, Elsevier, 52 , pp.675–694, 2008.

[26] Z. Zhang, S. Chen, and M. Yoon, "March: A distributed incentive scheme for peer-to-peer networks," Proc. of IEEE INFOCOM'07, May 2007.

[27] E. Damiani et al., "A reputation-based approach for choosing reliable resources in peer-to-peer networks," Proc. 9[th] ACM Conf. Computer and Comm. Security, ACM Press, pp. 207–216, 2002.

[28] Sabu M. Thampi and Chandra Sekaran K, "Autonomous data replication using Q-Learning for unstructured P2P networks," Proc. of the sixth IEEE Int'l Symposium on Network Computing and Applications (IEEE NCA07), , Cambridge, MA USA, ISBN: 0-7695-2922-4, pp. 311-317, July 12- 14, 2007.

[29] Sabu M. Thampi and Chandra Sekaran K, "Review of replication schemes for unstructured P2P networks", Proceedings of IEEE Int'l Advance Computing Conference IEEE (IACC'09), Thapar University, Patiala, ISBN: 978-981-08-2465-5, pp. 794-800, 2009.

[30] Sabu M. Thampi and Chandra Sekaran K, "Q-learning based collaborative load balancing using distributed search for unstructured P2P networks", Proc. of 33rd IEEE Conference on Local Computer Networks- LCN 2008, Montreal, pp. 797-802, 14-17 Oct. 2008.

[31] Yoshihiro Gotou, "Replication methods for enhancing search performance in peer-to-peer services," Master's Thesis, Department of Informatics and Mathematical Science, Graduate School of Engineering Science, Osaka University.